\documentstyle[aps,twocolumn]{revtex}

\begin{document}
\title{The spectral form factor is not self-averaging}
\author{R. E. Prange\cite{Md}}
\address{Laboratoire de Physique Quantique\cite{PS}, Universit\'e Paul Sabatier,\\
Toulouse, France}
\date{June 25, 1996}
\maketitle

\begin{abstract}
The spectral form factor, $k(t)$, is the Fourier transform of the two level
correlation function $C(x)$, which is the averaged probability for finding
two energy levels spaced $x$ mean level spacings apart. The average is over
a piece of the spectrum of width $W$ in the neighborhood of energy $\epsilon
_0$. An additional ensemble average is traditionally carried out, as in
random matrix theory. Recently a theoretical calculation of $k(t)$ for a 
{\em particular} system, with an energy average only, found interesting
nonuniversal semiclassical effects at times $t$ $\sim 1$ in units of $\tau
_\hbar =2\pi \hbar /(mean\;level\;spacing).$ This is of great interest {\em %
if} $k(t)$ is {\em self-averaging}, i.e, if the properties of a typical
member of the ensemble are the same as the ensemble average properties. We
here argue that this is not always the case, and that for many important
systems an ensemble average is essential to see detailed properties of $%
k(t). $ In other systems, notably the Riemann zeta function, it is likely
possible to see the properties by an analysis of the spectrum.
\end{abstract}


Recent work\cite{AAA} connects the energy level statistics of a {\em %
particular} {\em system}, say a Bunimovich stadium billiard, with the
predictions of {\em random matrix theory} [RMT]\cite{mehta}. This work has
generated great interest and some controversy. In particular it was
suggested that the form factor $k(t)$ of the spectrum of a given system can
be calculated theoretically to good approximation directly in terms of
certain classical parameters [i.e. the periodic orbits] of the system. For
scaling systems like billiards, not only does {\em the form factor approach
the universal RMT result} in an appropriate high energy limit, but
predictions\cite{AAA},\cite{BK} are made as to {\em how} this limit is
approached. These predictions differ in detail from one another, but their
gross features are the same.

It would obviously be of interest to take the spectrum of some system,
obtained either numerically or experimentally, and use it to calculate $k(t)$%
, so that comparison of the approximate theories with the numerically
`exact' structure factor can be made. The {\bf main result} of this paper is
that {\em in many cases }such a calculation is {\em unexpectedly difficult},
and may even be {\em impossible in principle}. We refer to this difficulty
as the fact that{\em \ the form factor is not a self-averaging quantity. }

Hard chaotic billiards without time reversal symmetry, `GUE'\cite{mehta},
the case worked out in detail up to now, unfortunately fall into the
probably impossible category. Billiards with nonisolated orbits are likely
to be difficult although not impossible to study numerically. However, the
Riemann zeta function, on the hypothesis that its zeroes are eigenenergies
of some linear Hermitean operator, can probably be studied effectively,
especially in view of the large database of zeroes obtained by Odlysko\cite
{Odlysko}.

Of all the two level correlation functions, [which are interrelated by
linear transformations], the form factor is of particular interest because,
according to the above ideas [which originated with Berry\cite{BFF}], the
dimensionless time $t$, the argument of $k(t)$, is essentially the relevant
classical orbit period in units of the Heisenberg time, $\tau _\hbar =2\pi
\hbar /\Delta $, where $\Delta $ is the mean level spacing. In particular,
for small times $t,$ it has long been established [and exploited\cite
{LenSpect}] that the form factor is strongly peaked at the periods of the
short periodic orbits, and the weights of these peaks are determined by the
stability properties of the orbits. Other correlators smear over the time
and are not so directly interpreted classically.

The {\em new feature }of the above theories is that there are definite{\em \
nonuniversal effects} in a {\em particular range} of $t$, namely near $t=1$.
These effects {\em disappear }in the high energy limit, as some power of
energy. Exactly {\em which} power determines whether the effects can be
observed in numerical calculations.

The form factor is usually loosely defined as the Fourier transform of the
two level correlation function $C(x)$, 
\begin{equation}
C(x)=\Delta ^2\left\langle \sum_{a,b}\delta (E+%
{\textstyle {1 \over 2}}
x\Delta -E_a)\delta (E-%
{\textstyle {1 \over 2}}
x\Delta -E_b)\right\rangle _W  \label{C(x)}
\end{equation}
where the spectrum is given by the sequence of levels $E_a.$ The average of $%
E$ is over an energy range $W$ about a central energy $\epsilon _0.$ An
additional average historically has been over an ensemble of systems, for
example ones given by RMT or by a random scattering potential. We are
investigating the consequences of disposing of this second average. The form
factor is then 
\begin{equation}
k(t)=\int dx\,e^{2\pi ixt}C(x).  \label{k-def}
\end{equation}

Two quite different techniques have been used. One\cite{AAA} starts from a
supersymmetric integral representation of $C(x)$, and then makes some
approximations which are discussed and justified. The other\cite{BK} extends
the diagonal approximation \cite{BFF} in an imaginative way to all values of 
$t$. In both cases it is assumed that the energy average{\em \ by itself }
validates the operation used to give the diagonal approximation, [DA], [for $%
t<<1$] which is equivalent to the idea that the actions of all orbits
[unrelated by symmetry] may be separately averaged. In addition, statistics
of classical orbits are invoked in the form of the Hannay-Ozorio de Almeida
sum rule\cite{HanOA}, and possible extensions of it\cite{AAA}.

A third technique\cite{QCA-RMT} explicitly combines random matrix theory\cite
{mehta} with orbital information about a particular system. It therefore 
{\em does} make an ensemble average, something like an average over the
random class of billiards all with the same area and also whose short orbits
have the same classical periods and stabilities. The results are less
complete, but do agree qualitatively with the other methods.

Ensemble averages are often used as a calculational device, even when no
ensemble is actually contemplated. The resistance of a particular copper
wire is conveniently calculated by averaging over an ensemble representing
the `random' position of impurities. An experimentalist can check the theory
by measuring the resistance of just one wire, however. That's because
resistance properties of a typical member of the ensemble are the same as
the average resistance over the ensemble: resistance is in this case a {\em %
self-averaging} quantity. Self-averaging is similar to {\em ergodicity.} For
an ergodic system, a time-averaged quantity is self-averaging.

Clearly self-averaging is a valuable trait. It is therefore disappointing to
find that the spectral form factor, by one standard definition, is {\em not
self-averaging.} Other definitions can be used, and some of them are
self-averaging. However, the alternative definitions known to us are
defective in the sense that in important cases they eliminate the predicted
effects: they are not sufficiently flexible to extract the small predicted
effects from the spectral data.

The energy average is essential, [even with an ensemble average], but we are
interested in the case that the results depend only weakly on the averaging
window $W$, and indeed, they should not depend much on exactly how the
average is made. This is supposed to be the case if the following two facts
hold.

\begin{itemize}
\item  There are a large number of energy levels in the average, i.e. $%
w\equiv W/\Delta >>1.$

\item  The averaging width $W$ is classically small. This means that energy
dependent objects like the mean level spacing, and other quantities
characterizing the underlying classical system, do not vary much, [or can be
linearized], in the energy range $\epsilon _0\pm \frac 12W.$ For short we
describe this as $W<<\epsilon _0,$ which tacitly supposes a natural zero of
the energy.
\end{itemize}

Theorists like to clean things up by taking a limit: $\epsilon
_0/W\rightarrow \infty ,$ $W/\Delta \rightarrow \infty .$ This requires a
system with an infinite number of levels, and also requires knowledge of how
things change as $\epsilon _0$ gets very large. A prominent example of such
a system is that of billiards. The order of limits is not completely trivial
however and another parameter is needed. Such limits should be left to the
very end, and then scrutinized carefully. In fact, numerical calculations
can't take such limits, so one must study how they deviate from the limiting
behavior.

The energy average alone certainly suffices to give interesting results for
small $t,$ $t\simeq t_c,$ where $t_c$ is the scale of the periods of the
shortest periodic orbits measured in units of the Planck time. For such
short times, $k\left( t\right) $ gives the {\em length spectrum}\cite
{LenSpect}$,$ that is, it has peaks at $t=T_p/\tau _\hbar $, i.e. at the
periods of the short periodic orbits, which for billiards are proportional
to the length of the orbits.

Here is a more careful definition. Let the energy average be given by a
Gaussian, i.e. $\left\langle f(E)\right\rangle _W=\frac 2{\sqrt{\pi }W}\int
dE\,f(E)\exp \left( -4(E-\epsilon _0)^2/W^2\right) .$ We argue next that the 
$x$ integral in Eq.(\ref{k-def}) should be cut off, and we choose a
Gaussian: $\exp (-(x\Delta /W_x)^2).$ We claim that a natural choice is $%
W_x\sim W$, and the particular choice $W_x=W$ is especially congenial.

Clearly $W_x$ should not be too large, since approximations relying on $%
x\Delta <<\epsilon _0$ must be made. If $W$ is as large as possible, then $%
W_{x\text{ }}$can't be much bigger. There is no advantage to taking $W$
unnecessarily small, as one would be throwing away data, in a numerical
calculation, for no gain. So, we take $W_x\leq W.$

The choice $W_x<<W$ has virtue. However, the same effect can be achieved by
making a time-convolution on $k(t),$ i.e. smearing each $t\,$ by an amount $%
\tau $. We find it convenient to choose this route. Therefore, we define 
\begin{equation}
k(t)=\int_{-\infty }^\infty dxe^{2\pi ixt}e^{-\frac{x^2}{w^2}}C(x)
\label{k_def2}
\end{equation}
where $C$ is based on the Gaussian energy average.

This $k(t)$ [except for $\left| t\right| <<1$] does {\em not} look like the
ensemble average. Rather, it has very considerable `noise' around the
ensemble mean value.

To see this, we remark that $k(t)\,$ can be written 
\begin{equation}
k(t)=\left| \sum_aF_ae^{2\pi i\frac{E_a}\Delta t}\right| ^2  \label{k(t)}
\end{equation}
where $F_a^2=\frac{2\Delta }{\sqrt{\pi }W}\exp (-4(E_a-\epsilon _0)^2/W^2)$.
The particular choice of cutoff $W_x=W$ has the advantage that the absolute
square of a single sum appears, rather than requiring a double sum. It
indicates, for example, that $k(t)$ should be close to a positive function,
no matter the form of the averaging. More general functions, $F_a,$
appropriately normalized, will do just as well. Note that for large $t$, one
expects just the diagonal terms in the double sum to survive, so that in
some sense $k\rightarrow 1$ in this limit. At very small $t$, it is expected
that $k(t)\simeq \delta (t)$. But in this case, the sum can be replaced by
an integral, and a Gaussian of width $\Delta /W$ replaces the $\delta $
function.

Eq.(\ref{k(t)}) makes it clear that for the spectrum of a chaotic system, $%
k(t)$, for large $t,$ is approximately given by a {\em random walk} in the
complex plane, of about $w$ steps. For the kind of systems envisaged, there
will be significant random walk character even for $t\sim 1.$ Insofar as the
phases of the individual terms in the sum Eq.(\ref{k(t)})\thinspace are
indeed independent random variables, one may show that the distribution of
values of $y=k(t)$ is 
\begin{equation}
\rho (y)=e^{-y}.
\label{rho0}
\end{equation}
This result is independent of the form of $F_a$ and of $W$, as long as there
are many terms in the sums of Eq.(\ref{k(t)}), i.e. as long as $w>>1$.
Basically, the result is an application of the central limit theorem. The
mean value [over an ensemble of such random variables] is $\left\langle
k(t)\right\rangle =1,$ and the variance of the distribution is $\left\langle
\left( k(t)-1\right) ^2\right\rangle =1.$

For smaller $t$, think of the $E_a$'s as ordered. Then it will take some
number of steps, of order $g(t),$ say, before the knowledge of the original
phase is lost. There will then be $w/g(t)$ effective random walk steps. The
effective step length is still of order unity, because the phase changes
systematically by an amount $e^{2\pi it\text{ }}$at each step. This gives a
formula for the distribution of $k(t)$, $\rho (k(t))=g(t)e^{-k(t)g(t)}.$ A
natural guess for $g(t)$ gives 
\begin{equation}
\rho (k(t))=\frac 1{k_E(t)}e^{-k(t)/k_E(t)},  \label{rho}
\end{equation}
where $k_E(t)$ is the ensemble average form factor.

Thus, for any averaging scale $W,$ the calculated value of $k(t)$ will
suffer large fluctuations, when $t$ is of order unity or greater. Increasing 
$W$ does not change the distribution, but it does make $k(t)$ vary more
rapidly. It's clear that if $t$ is changed by an amount of order $\Delta /W$
then $k(t)$ changes appreciably. For example, one estimates the random walk
average of $\left\langle \left| \frac d{dt}\sum_aF_ae^{2\pi it\left(
E_a-\epsilon _0\right) /\Delta }\right| ^2\right\rangle =\frac{4\pi ^2}{%
\Delta ^2}\sum_a\left( E_a-\epsilon _0\right) ^2F_a^2\sim \left( \pi
w\right) ^2$.

From this it follows that smearing the time $t$ by an amount $\tau $ is like
averaging over a number $\tau W/\Delta $ independent choices from the
distribution $\rho (k(t))$. Thus the `noise' in the smeared function $k_\tau
(t)$ is reduced by a factor $\sqrt{1/w\tau }$ and of course $k_\tau (t)$
changes appreciably only when $t$ is varied by amount of order $\tau $.

There is literature which numerically displays the noise and time averages
in $k(t)$ just discussed. A recent Letter\cite{EckMain} studies the spectrum
of hydrogen in a magnetic field. The distribution $\rho (y)$ is not
mentioned, although the scaling suggested in Eq.(\ref{rho}) is tacitly used
in taking the time average, [over $\tau \sim 0.6$]. The results agree
qualitatively with the discussion we have presented. There is a calculation%
\cite{Bl-Sm} of the statistics of the eigenphases of an $N\times N$
scattering matrix $S$, for a particular quantum chaotic scattering. The
function corresponding to $k(t)$ is $N^{-1}\left| 
\mathop{\rm Tr}
S^{tN}\right| ^2$ where $t$ is a discrete time, $t=n/N,$ and $t=1$ is the
Heisenberg time. In this case there is an ensemble of such systems which was
used to help to reduce the noise.

In the limit $\tau w\rightarrow \infty $ the noise disappears, and $k_\tau
(t)$ in that limit becomes self-averaging. For some purposes this is
adequate. For example Pandey\cite{Pandey}, `proved' that `all' spectral
correlation functions of particular members of an ensemble of random
matrices (and thus presumably $k(t)$), are self-averaging [or `ergodic']. Of
course, $C(x)$ as defined in Eq.(\ref{C(x)}) is manifestly {\em not }%
self-averaging, since it consists of lots of $\delta $-functions, while the
ensemble average is smooth. But Pandey argued that ``observable'' quantities
involve an integral over $x.$ Presumably the limits of such an integral are
fixed and independent of $\epsilon _0.$ This is energy smearing rather than
time smearing. Then in the limit $\epsilon _0\rightarrow \infty $ followed
by $W\rightarrow \infty $, Pandey indeed finds the energy smeared $%
C(x)\rightarrow $ random matrix prediction. Now one could, if desired,
define $k(t)$ as the Fourier transform of this $C(x)$. But since the large $%
W $ limit is taken first, for fixed finite $x$, it's the same as doing a
time average with $\tau W/\Delta \rightarrow \infty .$

We are more ambitious. Based on references \cite{AAA},\cite{BK},\cite
{QCA-RMT}, we predict for GUE billiards, that $\left| \left\langle
k(t)\right\rangle -k_R(t)\right| \sim 1/\sqrt{\epsilon _0}$, for $\left|
t-1\right| \sim 1/\sqrt{\epsilon _0}$. An ensemble average of $k$ is
indicated and $k_R$ is the random matrix form factor. Here units are chosen
so that $\Delta =1$, $mass=1/2$, and $\hbar =1.$

We therefore consider the `renormalized' function $h(s)\equiv \sqrt{\epsilon
_0}[k_\tau (1+s/\sqrt{\epsilon _0})-k_R(1+s/\sqrt{\epsilon _0})]$ which
should approach a theoretically predictable limit, `the signal', of order
unity, as $\epsilon _0\rightarrow \infty $ {\em provided} we can choose a
time smearing $\tau _{\text{ }}$which leads to a $k_\tau (t)$ sufficiently
close to the ensemble average prediction.

Put $w\simeq \epsilon _0^\beta ,$ and $\tau \simeq \epsilon _0^{-\alpha }.$
Clearly $\beta \leq 1$, in order that $W<<\epsilon _0$ for large $\epsilon
_0.$ If the time smearing is not to lose the signal, then $\alpha \geq \frac 
12.$ The `noise' in this case scales as $\epsilon _0^{(\alpha -\beta )/2}.$
Multiplying by $\sqrt{\epsilon _0}$, in order to calculate the noise in the
function $h$, gives a noise $\epsilon _0^\gamma $ where $\gamma \geq \frac 14%
.$ Thus, we predict that the noise {\em grows} at high energy, and that the
signal will be swamped.

Perhaps more sophisticated data-analysis schemes could extract the signal
from the noise, but all those we have tried were unsuccessful. One has to
keep in mind that this is not true noise: for a given spectrum it is
reproducible. Because, the noise grows at high energy in the renormalized $%
h(s),$ it hurts rather than helps to use very high energy data. In favorable
cases, it might be possible to use {\em low }energy data, but that does not
seem systematically possible. Thus, we characterize this situation as giving
unexpected numerical difficulties and perhaps it is impossible in principle
to check the theoretical predictions, {\em absent an actual ensemble average.%
}

Fortunately, there are systems which have much larger `quasiclassical'
corrections to the random matrix results, than do strongly chaotic
billiards. Most notable is the spectrum of nontrivial zeroes of the Riemann
zeta function. Although the hypothetical `Hamiltonian' whose spectrum
coincides with the zeroes is not known, or even proved to exist, the
`quasiclassical' parameters are known\cite{Odlysko}. In the language used
above, with $\epsilon _0$ having the mathematical meaning of height in the
complex plane along the critical line, the mean level spacing shrinks
according to $\Delta =2\pi /\ln (\epsilon _0/2\pi ),$ which we denote $%
\Delta \sim \epsilon _0^{-0}.$ The `orbit' periods in units of the Planck
time are $t_{p,m}=(m\Delta /2\pi )\ln p$, where $p$ is prime and $m$
repetitions of the fundamental orbit are taken. The stability weights, $a_p$
in the Gutzwiller formula, are $e^{-m\ln p/2}.$ We expect, based on the
techniques of reference\cite{QCA-RMT}, that corrections of magnitude $%
\epsilon _0^{-0}$ over a width $\left| t-1\right| \sim \epsilon _0^{-0}$
will exist. There should be enough data available\cite{Odlysko} to verify
this prediction.

Similarly, we think that GUE billiards with strong, nonisolated orbits, will
also have effects near $t=1$ observable in principle, but they are probably
too difficult in practice. The weight of a nonisolated orbit in the
Gutzwiller formula goes as $\epsilon _0^{-1/4}$ rather than $\epsilon
_0^{-1/2}$ as for unstable orbits. We expect corrections of magnitude $%
\epsilon _0^{-1/4}$ over a time window $\epsilon _0^{-1/4}$ thus allowing $%
\tau =\epsilon _0^{-1/4}.$ This leads to noise of magnitude $\epsilon
_0^{-1/8}$ in the rescaled $h(s).\,$Thus, with many energy levels in the
neighborhood of the $10^8$ level, such an effect might be seen in a direct
calculation of $k(t)$ from the spectrum. In general, a signal of width and
strength $\epsilon _0^{-\eta }$ can in principle be extracted from the data
on a single system only if $\eta <\frac 13.$

The much studied GOE class of systems (with time reversal symmetry)
unfortunately has very small effects at the Heisenberg time and should be
even harder to observe. However, GSE systems, (with sympletic symmetry)
might be favorable because the RMT result is singular at $t=1$ and thus
could suffer large corrections from the nonuniversal effects. These two
cases have not been worked out in any detail up to now, however.

Another question is whether it is possible to do ensemble averages of the
type mentioned, e.g. for a class of billiards. This might work if one has an
analog computer to do the job, namely an experimental system depending on a
parameter in a nice way, which can be changed sufficiently to give a further
average. Note that such a system is presented in reference\cite{Bl-Sm}.

We therefore conclude that by manipulations made on the spectrum of a
specific hard chaos billiard system, it is not known how, and it is perhaps
impossible, to confirm or discredit the theoretical predictions of $k(t)$
for $t\sim 1.$ For other systems, particularly the Riemann zeta function, it
is likely possible. The systems are distinguished by the order of magnitude
of the nonuniversal effects and their dependence on energy.

Reference\cite{AAA} above made no explicit ensemble or time averages, but
did not find a wide distribution for $k(t)$, but rather something like the
ensemble average of reference\cite{QCA-RMT}. Evidently an approximation
based on $W_x/W<<1,$ was made which discarded the noise, so their results
are correct if the signal has an energy exponent $\eta <\frac 13.$ It's less
clear what reference \cite{BK} predicts with respect to noise. Their result
for $k(t)$ is a sum of $\delta $-functions, $\delta (t-t_p)$ at the orbit
periods, but [we argue], these $\delta $-functions should have a width $%
\Delta /W.$ Except for $t$ very small, the spacing between the $t_p$'s will
be much smaller than this, since the spacing decreases exponentially with
increasing $t_p.$ [This is the exponential proliferation of orbits in
chaotic systems.] It was argued\cite{BFF} that in this case, the
Hannay-Ozorio de Almeida sum rule can be invoked with a smooth, non-noisy
result. If that argument is correct and carries over to the extensions of
the DA used, then these methods\cite{BK} do not produce the noise. If that
is so, the noise would have to come from the neglected off-diagonal
contributions. This might well be the case, as it is known that the
off-diagonal contributions are very important\cite{offdiag}. The `noise' in
the Bogomolny-Keating formulation thus presents a deep question, which will
probably not be answered in the near future.

Certain often used linear transforms of $k(t)$ {\em are} self-averaging, in
agreement with reference\cite{Pandey}. These are $\Sigma ^2(L)$ and $\Delta
_3(L),$ the number variance and the Dyson-Mehta spectral rigidity. The
former is given by $\Sigma ^2(L)=\int dt\,k(t)[\sin \pi Lt/\pi t]^2.$ For $%
L\sim 1$ or smaller, there is an effective time average over a scale of
order unity, thus eliminating the noise for large $W$. For large $L$, the
contribution of the integral comes from small $t,$ where the nonuniversal
signal dominates $k(t).$

Valuable discussions with O. Agam, E. Bogomolny, O. Bohigas, S. Fishman, B.
Georgeot, J. Keating, D. Poilblanc, and C. Sire are gratefully acknowledged.
We thank J. Bellissard for hospitality at the Universit\'e Paul Sabatier.

\end{document}